\documentclass[prd,12pt,onecolumn,floatfix,superscriptaddress]{revtex4-1}
\usepackage{dcolumn,amsmath}
\usepackage{amssymb}
\usepackage{graphicx}
\usepackage{bm}
\usepackage{amssymb}
\usepackage{multirow}

\usepackage{amsfonts}
\usepackage{amsmath}
\usepackage{cancel}
\usepackage{xcolor}

\begin{document}


\title{$\mathcal{P}$, $\mathcal{T}$-violating axion-mediated interactions in RaOH molecule}

\author{Anna Zakharova}\email{zakharova.annet@gmail.com, anna.zakharova@spbu.ru} 
\affiliation{St. Petersburg State University, St. Petersburg, 7/9 Universitetskaya nab., 199034, Russia}
\affiliation{Petersburg Nuclear Physics Institute named by B.P. Konstantinov of National Research Centre
"Kurchatov Institute", Gatchina, 1, mkr. Orlova roshcha, 188300, Russia}

\date{Received: date / Revised version: date}
\begin{abstract}
If axion simultaneously has the scalar couplings to the nucleons and pseudo-scalar couplings to the electrons,
it may mediate a $\mathcal{P}$, $\mathcal{T}$-violating interaction between the electronic shell and nuclei in the
molecules. The polyatomic molecule RaOH, which is considered as a promising platform for the $\mathcal{P}$, $\mathcal{T}$-violation searches, is studied for its sensitivity to such interactions. Due to the long-range nature (on molecular scales)
of the axion-mediated interaction, it is important whether the molecular parameter would be sensitive to the
vibration of the molecule.
Our results imply that the impact of the vibrations on the axion-mediated electron-nucleon interaction in the molecule is
similar to the impact on the short-range electron-nucleon scalar-pseudoscalar interaction studied earlier.
\end{abstract}
\maketitle
\section{Introduction}
\label{intro}

An important question in fundamental physics is the origin of Dark matter, the existence of which is predicted from the rotational curves of galaxies, and data from gravitational lensing of galaxy objects, such as the Bullet cluster. It also allows us to describe the formation of structures in the early Universe, and the observed inhomogeneities in cosmic microwave background radiation \cite{chadha2022axion}.

Pseudoscalar particles called axions are promising potential candidates for Dark matter constituents \cite{cosmo:1983, abbott1983cosmological,dine1983not}. Such particles with certain masses and coupling constant values may solve the strong $\mathcal{C}$$\mathcal{P}$-problem \cite{peccei1977cp}, which remains an open problem of the Standard Model (SM) of particle physics \cite{schwartz2014quantum,ParticleDataGroup:2024cfk}. QCD does not have other sources of $\mathcal{C}$$\mathcal{P}$-violation except the axion-associated $\theta$-term. 
In the SM, all known parity violations occur in the weak interaction, related to the mixing of the quark matrices -- Cabibbo-Kobayashi-Maskawa \cite{Cabibbo1963,KobayashiMaskawa1973}, and the neutrino matrices -- Pontecorvo-Maki-Nakagawa-Sakata \cite{Pontecorvo1957,MNS1962}. Some physical effects, such as the electron electric dipole moment (eEDM), and scalar-pseudoscalar electron-nucleon interaction (NE-SPS), are strongly suppressed in the Standard Model \cite{Fukuyama2012,PospelovRitz2014,YamaguchiYamanaka2020,YamaguchiYamanaka2021}. However, some $\mathcal{CP}$-violating scenarios, such as SUSY, predict much larger values for these effects. The existence of new sources of $\mathcal{CP}$-violation may also be necessary to explain the problem of baryon asymmetry of the Universe, and may lead to a value of $\theta$-term closer to the neutron electric dipole moment limit \cite{sakharov1998violation}.

We also can consider axion-like particles (ALPs) that are light pseudoscalar particles that may have a qualitatively similar behavior to the QCD axion but do not necessarily address the strong CP problem. Such particles may appear in diverse scenarios, resulting in a variety of their properties.

Molecular experiments allow us to study the effects related to $\mathcal{P}$ and $\mathcal{T}$-parity violation \cite{ginges2004violations,baron2014order, PospelovRitz2014,ChubukovLabzowsky2016,ACME:18}. For example, an eEDM limit was obtained in an experiment with the diatomic molecule HfF$^+$ \cite{roussy2023improved}. Theoretical calculations are necessary to estimate the parameters of $\mathcal{P}$ and $\mathcal{T}$-parity violation for these effects \cite{KozlovLabzowsky1995}. The phenomena that may manifest themselves in the molecular spectra include the interaction with  cosmic fields of axions and Dark photons, most straightforwardly in the chiral species such as the ground and vibrational states of the CHBrClF molecule \cite{gaul2020chiral}. Such interaction was also considered for a non-chiral molecule $RaOCH_3$in non-equilibrium configurations \cite{zakharova2025hexatomic}  

Among the studies of the effects caused by new interactions with axions in molecules, one can highlight the works on the effects of axion exchange between electrons and nuclei, as well as between electrons and electrons, studied for diatomic molecules HfF$^+$ \cite{Prosnyak:2023duq}, BaF \cite{prosnyak2024axion}, and the triatomic molecule YbOH \cite{maison2020study, maison2021axion}. Also, the interaction mediated by the axion produced by an electron-axion pseudoscalar vertex and a nucleon-axion or electron-axion scalar vertex was considered in \cite{stadnik2018improved,dzuba2018new}. In these works, the contribution of such an interaction to the dipole moment of atoms and molecules was studied, and constraints on the corresponding interaction constants were obtained.

The polyatomic molecules such as the triatomic molecules RaOH and YbOH are a promising platform for the future parity violation searches.
It is important to stress that the measurement of the $\mathcal{P}$, $\mathcal{T}$ on the triatomic molecules involves their excited rovibrational states. In addition to the doublets associated with the rotation of the electron shell, triatomic linear molecules have $l$-doublets \cite{ourRaOH, zakharova2021rovibrational, zakharova2022impact,petrov2022sensitivity}. In hexatomic molecules of the "symmetric top" type, additional doublets appear associated with the rotation of the CH$_3$ ligand in two opposite directions \cite{zakharova2022rotating,zakharova2024symmetric,zakharova2025hexatomic}. Measurement of the relative shifts within different parity doublets allows one to suppress significantly many systematic effects. The possibility of laser cooling of the polyatomic molecules also presents a huge advantage for the experiment \cite{isaev2016polyatomic, kozyryev2016proposal,kozyryev2019determination,augenbraun2021observation, mitra2020direct,fan2021optical}.
The laser cooling for the triatomic molecule SrOH was demonstrated in \cite{sawaoka2025optical}. For the molecule RaOH only limited spectroscopic data exists, such as in \cite{conn2025production} where the band origins, rotational constant and spin-rotation constant was obtained.

The axion-mediated interaction can manifest itself in the molecular spectra the same way as such P, T violating effects as the electron EDM and the scalar-pseudoscalar electron-nucleon interaction. It is important to compute the sensitivities to the different effects to be able to distinguish them from each other using multiple measurements on different molecular species.

If we suppose the existence of the axion exchange interaction of the electrons with nuclei characterized by the product of coupling con\-stants $g_{e,P}\cdot g_{N,S}$, this $\mathcal{P}$, $\mathcal{T}$-odd effect can be estimated from the   splitting between levels with opposite values of $M$ (projection of total angular momentum on space-fixed frame) given by:
\begin{equation}
\Delta E_{\mathcal{P},\mathcal{T}}\simeq Pg_{e,P}g_{N,S}\Omega W_{\rm ax}  ,
\label{split}
\end{equation}
where $\Omega$ is the projection of the electron spin on the molecular axis.
The parameter $W_{\rm ax}$ is determined from the molecular electronic structure cal\-cu\-la\-tion. $P\leq 1$ is the degree of polarization of the molecule \cite{petrov2022sensitivity}.
Because the states of interest are excited rovibrational states, it is important to study the impact of the molecule geometry on the molecular parameters. Previously, we have studied such impact on the enhancement of the eEDM, and NE-SPS for the triatomic molecules RaOH \cite{ourRaOH}, YbOH \cite{zakharova2021rovibrational} as well as hexatomic molecules RaOCH$_3$ \cite{zakharova2022rotating} and YbOCH$_3$\cite{zakharova2024symmetric}. While the impact is noticeable, it is not particularly large for these properties. The axion-mediated interaction differs from these interactions because of its long-range nature, therefore, one may find it interesting to check whether the impact of the molecular vibrations may become significant in this case. In this paper we study the molecular parameter for the electron-nucleon axion-mediated interaction for the triatomic molecule RaOH.

\section{Axion-mediated interaction of electrons and nuclei}

The interaction terms between the axion and the fermions of the Standard model may be written as
\begin{equation}
\mathcal{S}_{\rm int} = \sum_{f\in\mathrm{SM}} \Big(g_{f,S}\bar{\Psi}_f a \Psi_f + i g_{f,P}\bar{\Psi}_f a\gamma_5\Psi_f\Big).
\end{equation}
\begin{figure}
    \centering
    \includegraphics[width=0.5\linewidth]{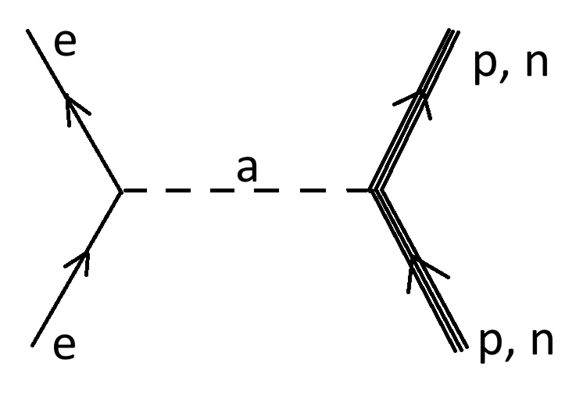}
    \caption{Feynman diagram for the axion-mediated electron-nucleon interaction}
    \label{fig:feynman}
\end{figure}

If both scalar interaction with nucleons $g_{N, S}$ (with $N=p, n$), and pseudo-scalar interaction with electrons $g_{e, P}$ are present, the process of axion exchange represented by a Feynman diagram in Fig. \ref{fig:feynman} induces the following one-electron operator of electron interaction with the nucleus $A$, written in the atomic units as:
\begin{equation}
\hat{H}_{eN} =
i\frac{g_{e, P}}{4\pi}\sum_{i=1}^{N_e}\sum_{k=1}^{A}\int d^3\mathbf{R}_k\, g_{N_k, S}\rho_k(\mathbf{R}_k)\frac{e^{-m_ac|\mathbf{r}_i-\mathbf{R}_k|}}{|\mathbf{r}_i - \mathbf{R}_k|}\gamma_0^{(i)}\gamma_5^{(i)},
\end{equation}
where $N_e$ is the number of electrons, $A$ - is the atomic weight of the nucleus (i.e. number of nucleons), $\mathbf{r}_i$ is the coordinate of the $i$-th electron, $\mathbf{R}_k$ is the coordinate of the $k$-th nucleon, $N_k=p,n$ is the type of nucleon, $\rho_k$ is the nucleon density function normalized by $1$, $m_a$ is the axion mass. The gamma matrices $\gamma_0^{(i)}$ and $\gamma_5^{(i)}$ are acting on the spinor of the $i$-th electron. In the atomic units the speed of light is $c\simeq 137$.

A number of approximations can be employed in this operator.

First, we approximate the nucleon density with a single function $\rho$, which we can identify as the charge density.
\begin{equation}
\hat{H}_{eN} \simeq
i A \frac{g_{e,P}\bar{g}_{N, S}}{4\pi}\sum_{i=1}^{N_e} \int d^3\mathbf{R}\,\frac{\rho(\mathbf{R})e^{-m_ac|\mathbf{r}_i-\mathbf{R}_k|}}{|\mathbf{r}_i-\mathbf{R}|}\gamma_0^{(i)}\gamma_5^{(i)},
\end{equation}
where we introduced $\bar{g}_{N,S}$ for the nucleus with the atomic weight $A$, and charge $Z$:
\begin{equation}
    \bar{g}_{N,S}=\frac{(A-Z)g_{n, S} + Z g_{p, S}}{A}.
\end{equation} 
In fact, the results of \cite{prosnyak2024axion} imply that the non-point-like description of the atomic nuclei becomes important only for $m_a\gtrsim 10\,{\rm MeV}$. This is consistent with our own computation.

The molecular parameter is then introduced as
\begin{equation}
W_{\rm ax} = \frac{1}{g_{e,P}\bar{g}_{N,S}{\Omega}}\langle\Psi|\hat{H}_{eN}|\Psi\rangle,
\end{equation}
$\gamma_5$ presented in the operator requires electrons gaining relativistic momenta for this interaction to have any significance. Combined with $A$-scaling, this means that, despite its long-range nature, only the contribution of the interaction of electrons with heavy nuclei and their neighborhood (though significant in size) will matter. That's why for the RaOH molecule, considered in this paper, we take into account only the contribution of the interaction of electrons with the heavy  ($A=226$) Radium atom.
The Eq.(5) gives the expectation value of the axion-
mediated $\mathcal{P}$ and $\mathcal{T}$-violating part of Hamiltonian. This value can be used to obtain product of constants $g_{e,P}\cdot g_{N,S}$ from energy shifts of molecular levels that could, in principle, be detected experimentally.

Second, the typical molecular length scale corresponds to the Compton length of masses in $\sim 1\,{\rm KeV}$ range. This is much larger than the astrophysically acceptable masses of the axion. Therefore, we can approximate the exponent in the numerator as $e^{-m_a|\mathbf{r_i}-\mathbf{R}_k|}\simeq 1$. However, it is still interesting to consider the opposite limit of $m_ac\gtrsim 10\,{\rm GeV}$ in which this interaction becomes an effective NE-SPS interaction \cite{prosnyak2024axion}:

\begin{equation}
\hat{H}_{s}=
ig_{s}\frac{G_F}{\sqrt{2}}
Z\rho(\vec{r})\gamma_0\gamma_5
\end{equation}
with the parameter describing the sensitivity to this interaction:
\begin{equation}
W_s=\frac{1}{g_s\Omega}\langle\Psi|\hat{H}_s|\Psi\rangle,
\end{equation}
where the Fermi constant is $G_F = 2.2225\cdot 10^{-14}$ in the atomic units.
In the high-mass limit when $m_a c$ comparable and less to the radius of the nucleus, these parameters are related to each other as
\begin{equation}
W_{\rm ax}\simeq \frac{\sqrt{2}A}{Z G_F}\frac{W_{\rm s}}{m_a^2 c^2}
\label{WsApprox}
\end{equation}
As we have previously computed the value of $W_s$ in \cite{ourRaOH}, we will use this relation to check the validity of our computation.

\section{One-center restoration method}

The computational complexity for the RaOH molecule is high due to the large number of electrons and the necessity to introduce large number of the basis functions to account for the states with large momenta. To overcome that, we use generalized effective core potential (GRECP) to describe the core electrons \cite{titov1999generalized,mosyagin2010shape,mosyagin2016generalized}. However,  only the 2-component solutions can be obtained this way.  And they have incorrect smoothed behavior in the core region. To overcome that problem, the one-center restoration method (OCR) was developed in \cite{titov2006d}. Let us elucidate the main idea of this method.

Firstly, we construct the equivalent bases. The atomic problem for the heavy atom is solved using two approaches to get two sets of solutions. The first set of solutions is obtained using the full-electron Harthree-Fock-Dirac method, resulting in:
\begin{equation}
\Phi_{nljm}(\mathbf{r}) = \begin{pmatrix}f_{nlj}(r)\mathcal{Y}_{ljm}(\theta,\phi)\\i g_{nlj}(r)\mathcal{Y}_{2j-l,jm}(\theta,\phi)\end{pmatrix},
\end{equation}
where $\mathcal{Y}_{ljm}$ are 2-component spinor spherical harmonics. The second set of solutions is obtained using GRECP resulting in 2-component basis:
\begin{equation}
\tilde{\Phi}_{nljm}(\mathbf{r}) = \tilde{f}_{nlj}(r)\mathcal{Y}_{ljm}(\theta,\phi).
\end{equation}
These two sets of solutions are mapped to each other based on  an identification of the quantum numbers. This way the equivalent bases are constructed.

Second, the 2-component molecular spinor $\tilde{\Psi}$ obtained in the GRECP approximation is represented in the vicinity of the nucleus as the superposition of the 2-component basis:
\begin{equation}
\tilde{\Psi}(\mathbf{r})=\sum_{njl} C_{njlm}\tilde{\Phi}_{njlm}(\mathbf{r}),
\end{equation}
and then the 4-component spinor is constructed using the same coefficients but 4-component basis functions equivalent to the 2-component ones:
\begin{equation}
\Psi(\mathbf{r})=\sum_{njl} C_{njlm}\Phi_{njlm}(\mathbf{r}).
\end{equation}
The property is computed using the constructed 4-component molecular spinors. Details of implementation are given in \cite{zakharova2025parallelized}.

\section{Coupled-channels technique}

\begin{figure}[h]
\centering
  \includegraphics[width=0.25\textwidth]{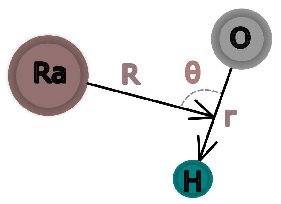}
  \caption{Jacobi coordinates}
  \label{Jacob}
\end{figure}

For the molecular wavefunction we assume the validity of the Born-Oppenheimer approximation:
\begin{equation}
\Psi_{total}\simeq\Psi_{nuc}(R, \hat{R}, \hat{r})\psi_{elec}(R,\theta|q),
\end{equation}
$q$ means  coordinates of the electronic subsystem,
$\hat{r}$ and $\hat{R}$ are directions of OH axis and Ra - center mass of OH axis respectively, $\theta$ is the angle between above axes, $R$ is the distance between the Radium atom and the ligand center of mass (Fig. \ref{Jacob}). We consider the fixed ligand stretch since the Ra -- OH bond stretching and bending frequencies are much smaller than the vibrations of the OH, using the equilibrium distance $r=1.832 \,a.u.$ \cite{Huber:1979}.
The Dirac-Coulomb equation in the classical field of fixed nuclei has a solution $\psi(R,\theta|q)$.
Corresponding Hamiltonian can be written in such form:
\begin{equation}
\hat{H}_{nuc}=-\frac{1}{2\mu}\frac{\partial^2}{\partial R^2}+\frac{\hat{L}^2}{2\mu R^2}+\frac{\hat{j}^2}{2\mu_{OH}r^2}+V(R,\theta),
\end{equation}
where $\mu$ and $\mu_{OH}$ are the reduced masses of the $Ra-OH$ and the OH ligand 
respectively, $\hat{L}$  is the angular momentum of the rotation of the heavy Radium atom and the ligand around their common center of mass, $\hat{l}$ is the angular momentum of the ligand rotation, and $V(R,\theta)$ is the effective adiabatic potential.

Thus, the solution of the Schr\"{o}dinger equation is the nuclear wavefunction $\Psi_{nuc}(R, \hat{R}, \hat{r})$: 
\begin{equation}
\hat{H}_{nuc}\Psi_{nuc}(R, \hat{R}, \hat{r}) = E \Psi_{nuc}(R, \hat{R}, \hat{r}).
\label{Shreq}
\end{equation}
We use the following series expansion to solve equation (\ref{Shreq}):
\begin{equation}
\Psi_{nuc}(R, \hat{R}, \hat{r}) = \sum_{L=0}^{L_{max}}\sum_{j=0}^{j_{max}} F_{JjL}(R)\Phi_{JjLM}(\hat{R},\hat{r}),
\label{psiexp_RaOH}
\end{equation}
where
\begin{equation}
\Phi_{JjLM}(\hat{R},\hat{r}) = \sum_{m_L,m_j} C^{JM}_{Lm_L,jm_j} Y_{Lm_L}(\hat{R})Y_{jm_j}(\hat{r}),
\end{equation}
 $Y_{Lm_L}$ is spherical function, $C^{JM}_{Lm_L,jm_j}$ are the Clebsch-Fock-Gordan coefficients.

Thus, we have a system of close-coupled equations for $F_{JjL}(R)$ \cite{mcguire1974quantum}:
\begin{equation}
\Big[\frac{d^2}{dR^2}-\frac{L(L+1)}{R^2}-\frac{\mu j(j+1)}{\mu_{OH} r^2}+2\mu E\Big]F_{JjL}(R)
=2\mu\sum_{\tilde{j},\tilde{l}}\mathcal{V}_{jL,\tilde{j}\tilde{L}}^J F_{J\tilde{j}\tilde{L}}(R),
\end{equation}

where on the right side we have introduced a matrix potential:
\begin{align}
\mathcal{V}_{jL,\tilde{j}\tilde{L}}^J(R)=(-1)^{J+j+\tilde{j}}\sqrt{(2L+1)(2\tilde{L}+1)(2j+1)(2\tilde{j}+1)}\nonumber\\
\times \sum_{k=0}^{k_{max}}V_{k}(R) \begin{Bmatrix}j&L&J\\\tilde{L}&\tilde{j}&\lambda\end{Bmatrix}
\begin{pmatrix}L&\lambda&\tilde{L}\\0&0&0\end{pmatrix}\begin{pmatrix}j&\lambda&\tilde{j}\\0&0&0\end{pmatrix}
\end{align}
where a matrix with curly brackets denotes a $6j$-symbol, matrices with round brackets denote $3j$-symbols, and $V_k(R)$ correspond to the decomposition of the potential in terms of the Legendre polynomials $P_k(x)$:
\begin{equation}
V(R,\theta) = \sum_{k=0}^{k_{max}}V_k(R) P_k(\cos\theta).
\end{equation}

To deal with the dependence on $R$ we use the following approach. We choose a basis $f_n(R)$ of the eigenfunctions for longitudinal oscillations in the harmonic approximation:
\begin{equation}
\Big[-\frac{1}{2\mu}\frac{d^2}{dR^2}+\frac{\mu\omega_\parallel^2}{2} (R-R_{eq})^2\Big] f_n(R) = \omega_\parallel\Big(n + \frac{1}{2}\Big)f_n(R).
\end{equation}

Then we use the following decomposition:
\begin{equation}
F_{JjL}(R)=\sum_{n=0}^{n_{max}} F_{JjLn} f_n(R).
\end{equation}
The exact solution would correspond to the infinite sum but for the numerical computations we use the finite basis approximation, and cut our sum to $n_{max}$. We considered different values of $n_{max}$ to check numerical convergence. Here $n_{max}$ is a maximal occupation number for the longitudinal vibrations. This means that higher value of $n_{max}$ describes wider wavefunctions as well as more rapid oscillations in $R$. The electronic computations are performed on a discrete grid of $R$ with minimal and maximal values defined $R_{min}$ and $R_{max}$ correspondingly.
We introduce matrix elements for the centrifugal term is:
\begin{equation}
C_{n\tilde{n}}=\int_{R_{min}}^{R_{max}} dR\,f_n(R)\frac{1}{R^2}f_{\tilde{n}}(R),
\end{equation}
and the potential:
\begin{equation}
\tilde{V}_{kn\tilde{n}}=\int_{R_{min}}^{R_{max}} dR\,f_n(R) \Big(V_k(R)-\delta_{k,0}\frac{\mu\omega_{\parallel}^2}{2}(R-R_{eq})^2\Big)f_{\tilde{n}}(R).\label{Integ}
\end{equation}
The values for $R_{min}$ and higher $R_{max}$ are chosen in such a way that the obtained nuclear motion wavefunctions decrease rapidly beyond that range and do not impact the accuracy of the numerical computation.

Then we get the finite linear system:
\begin{equation}
\Bigg[-2\mu\omega_\parallel\Big(n+\frac{1}{2}\Big)-\frac{\mu j(j+1)}{\mu_{OH} r^2}+2\mu E\Big]F_{JjLn}
=2\sum_{n}L(L+1)C_{n\tilde{n}}F_{JjL\tilde{n}} + 2\mu\sum_{\tilde{j},\tilde{l},\tilde{n}}\tilde{\mathcal{V}}_{jLn,\tilde{j}\tilde{L}\tilde{n}}^JF_{J\tilde{j}\tilde{L}\tilde{n}},\label{FiniteEq}
\end{equation}
where we use:
\begin{align}
\tilde{\mathcal{V}}_{jLn,\tilde{j}\tilde{L}\tilde{n}}^J=(-1)^{J+j+\tilde{j}}\sqrt{(2L+1)(2\tilde{L}+1)(2j+1)(2\tilde{j}+1)}\nonumber\\
\times \sum_{k=0}^{k_{max}}\tilde{V}_{kn\tilde{n}} \begin{Bmatrix}j&L&J\\\tilde{L}&\tilde{j}&\lambda\end{Bmatrix}
\begin{pmatrix}L&\lambda&\tilde{L}\\0&0&0\end{pmatrix}\begin{pmatrix}j&\lambda&\tilde{j}\\0&0&0\end{pmatrix}\label{PotMatEl}.
\end{align}

We use the following procedure to process the potential:
\begin{itemize} 
\item {Adiabatic potential values are obtained in the electronic computations for a discrete grid of ($R_i$, $\theta_j$)}
\item We fit values for $\theta=0^\circ$ using the Morse potential:
\begin{equation}
V(R, 0^\circ)=D_e \Bigg(1-\exp\Big(-A(R-R_{eq})^2\Big)\Bigg),
\end{equation}
which gives $R_{eq}$ and $\omega_\parallel = \sqrt{\frac{2D_e}{\mu}}A$.
\item{The dependence on $\theta$ is interpolated using Akima splines for each value of $R_i$.}
\item{The Gauss-Legendre method is then used to obtain the coefficients of the Legendre polynomial expansion $V_k(R_i)$ on the grid of $R_i$ values.}
\item {Then the dependence of the expansion coefficients $V_k$ on $R$ is also interpolated by Akima splines.}
\item{The resulting interpolation is used to calculate the integrals $V_{kn\tilde{n}}$, Eq.\eqref{Integ}}
\end{itemize}

Applying similar processing (except second step with Morse potential fit) and equation \eqref{PotMatEl} to the property values, we also obtain the matrix elements of the properties $W_{jLn,\tilde{j}\tilde{L}\tilde{n}}$. The property average is obtained as
\begin{equation}
\langle W_{ax}\rangle = \sum_{j,l,n}\sum_{\tilde{j},\tilde{l},\tilde{n}}F_{JjLn}^\ast F_{J\tilde{j}\tilde{L}\tilde{n}}W_{jLn,\tilde{j}\tilde{l}\tilde{n}}
\end{equation}

\section{Results and discussion}

The molecular orbitals were obtained in DIRAC19 software package using self-consistent field (SCF) level. To describe the heavy Radium atom we used a 10-valence electron basis set adapted to GRECP developed by the Quantum Physics and Chemistry Department of PNPI \cite{QCPNPI:Basis}. We have used the equivalent bases adapted to this potential previously used in \cite{ourRaOH}.

The functions $f_{nlj}$ and $g_{nlj}$ used to describe the equivalent bases functions are represented as a set of values on the uneven grid of $r_i$ starting with $r_1=7.04\cdot 10^{-6}{\rm a.u.}$ to $150 {\rm a.u.}$ Inside the nucleus region $r < r_1$ the following ansatz is used:
\begin{equation}
f_{nlj}(r) = r^\gamma\sum_{k=0}^5 a_k r^k.
\end{equation}
This six terms correspond to a special anzats to represent a wave function in the core region. This choice of anzatz was made before in the works on the one-center restoration technique \cite{titov2003accuracy} and used in the preceding work with same equivalent basis sets for Radium atom \cite{ourRaOH}. This is important for an accurate description of the short-range electron-nucleon interactions. However, in our case, because of the significant long-range contribution, this region contribution is negligible when $m_a\ll 1{\rm MeV}$. This is in sharp contrast to the eN-SPS interaction.

\begin{figure}
    \centering
    \includegraphics[width=0.7\linewidth]{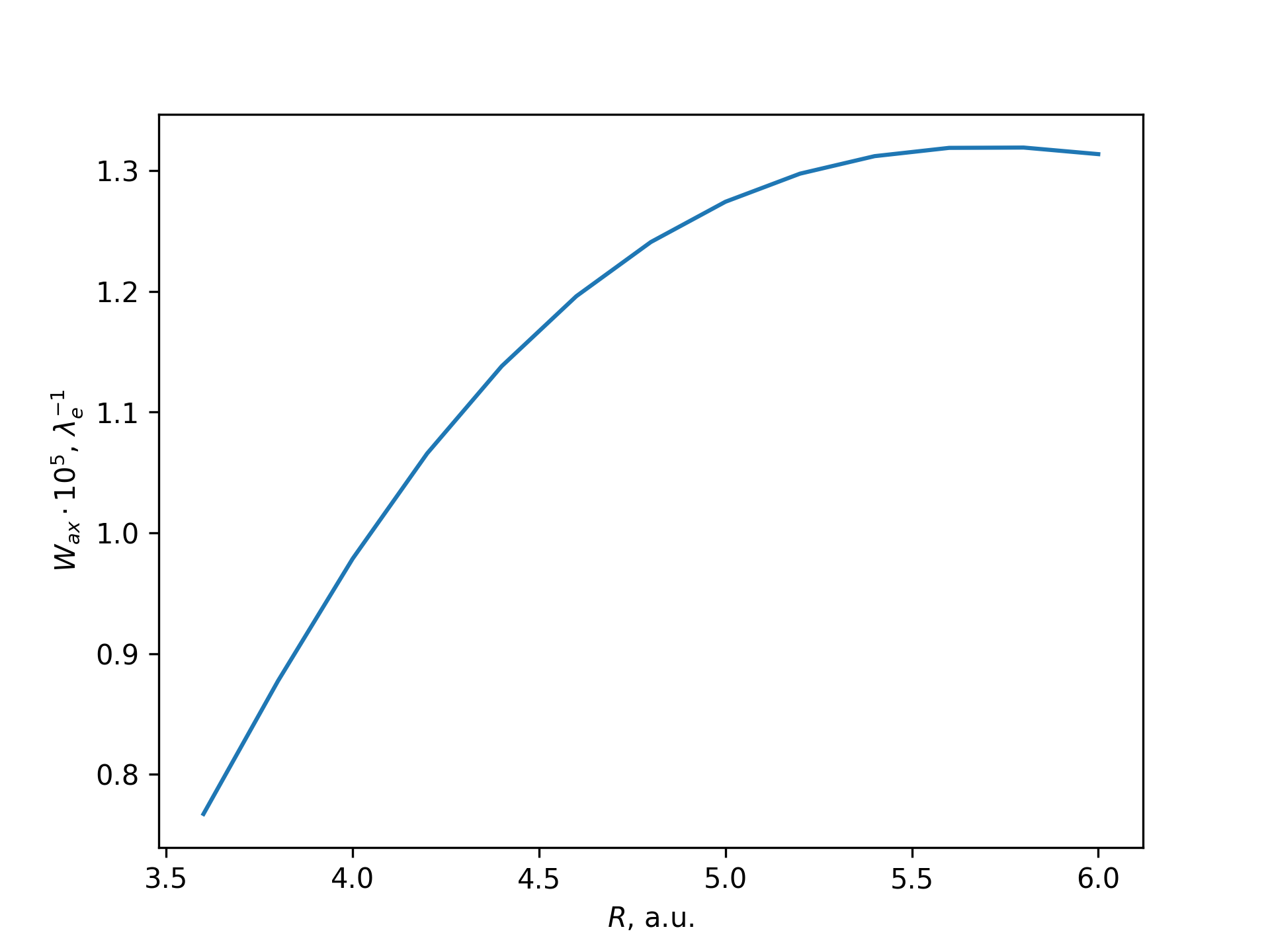}
    \caption{The dependence of $W_{ax}$ on $R$ for $\theta=0^\circ$}
    \label{fig:plot2D}
\end{figure}

\begin{figure}
    \centering
    \includegraphics[width=0.7\linewidth]{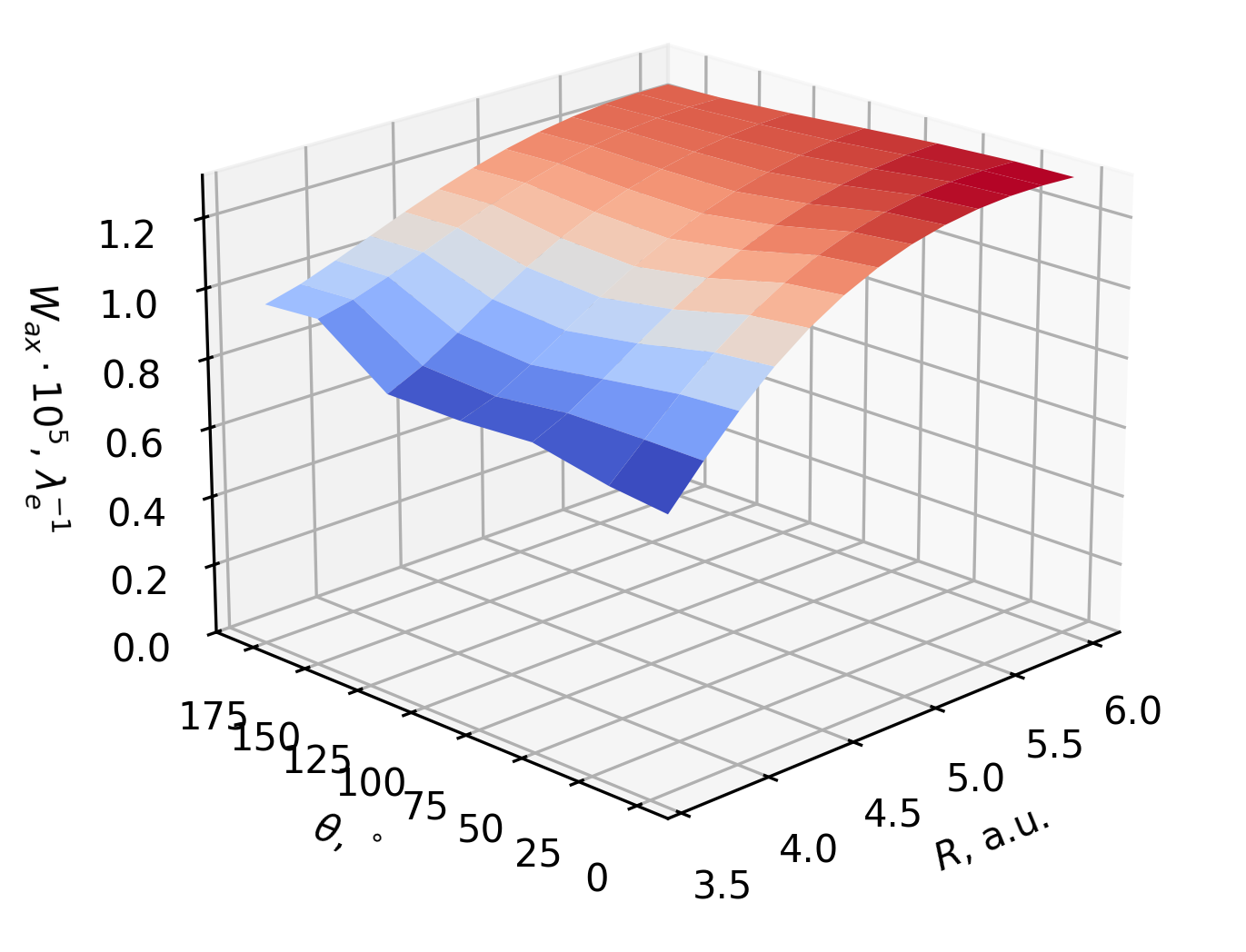}
    \caption{The dependence of $W_{ax}$ on $R$ and $\theta$}
    \label{fig:plot3D}
\end{figure}

The equilibrium geometry was determined in \cite{ourRaOH} to correspond to $R=4.4\,{\rm a.u.}$ The property was computed on a grid of values of $R$ and $\theta$. $R$ was taken to span values from $3.6\,{\rm a.u.}$ to $6.0\,{\rm a.u.}$ with step $0.2\,{\rm a.u.}$, where $\theta$ was taken to be $0^\circ$, $25^\circ$, $57^\circ$, $90^\circ$, $122^\circ$, $155^\circ$ and $180^\circ$. The resulting value for the equilibrium geometry equals $1.1407\cdot 10^{-5}\lambda_e^{-1}$ where $\lambda_e = \frac{\hbar}{m_ec}$ is the electron Compton length. Compare this with $3.36\cdot 10^{-5}\lambda_e^{-1}$ for YbOH obtained in \cite{maison2020study}. The dependence on the $\theta$ and $R$ is depicted on the Fig. \ref{fig:plot2D} and Fig. \ref{fig:plot3D}.  The values for several rovibrational states are given in Table \ref{tbl:table1}. To check the convergence and estimate numerical error the computation was performed for $l_{max}, j_{max}, k_{max} = 40,\ldots 43$, and $n_{max}=10,\ldots 13$.

\begin{table}[h]
\small
  \caption{Electron-nucleon axion-mediated interaction}
  \label{tbl:table1}
  \renewcommand{\arraystretch}{1.5}
  \begin{tabular*}{0.7\textwidth}{@{\extracolsep{\fill}}llllr}
    \hline\hline
$v_{\parallel}$ & $v_\perp$ &  $l$ &  $W_{ax}$,  $10^{-5}\lambda_e^{-1}$ & Numerical relative error\\
\hline
  \multicolumn{5}{l}{Equilibrium configuration}\\
\hline
 &  &  & 1.1407    & 1.6791 $\cdot 10^{-5}$\\
\hline
  \multicolumn{5}{l}{J=0}\\
\hline
0 & 0 & 0 & 1.1347    & 1.6966 $\cdot 10^{-7}$\\
1 & 0 & 0 & 1.1367 & 1.7668 $\cdot 10^{-7}$\\
2 & 0 & 0 & 1.1379 & 5.3008 $\cdot 10^{-7}$\\
0 & 2 & 0 & 1.1208 & 5.1133  $\cdot 10^{-7}$\\
\hline
  \multicolumn{5}{l}{J=1}\\
  \hline
0 & 0 & 0 & 1.1347 & 1.6966  $\cdot 10^{-7}$ \\

0 & 1 & 1 & 1.1278 & 3.7887 $\cdot 10^{-7}$\\

\hline\hline
  \multicolumn{5}{l}{The YbOH equilibrium (FS-CCSD, $m_a=1..10$eV) \cite{maison2020study}}\\
  \hline
&  &  & 3.36 &    \\
 \hline\hline
  \multicolumn{5}{l}{The BaF equilibrium  (CCSD, $m_a=1..10^2$eV) \cite{prosnyak2024axion}}\\
  \hline
&  &  & 1.77 &    \\
 \hline\hline
  \end{tabular*}
\end{table}


Our results show that, despite the long-range nature of the property in question, its dependence on the geometry of the molecule is very similar to the short-range eN-SPS interaction. However, it should be stressed that previous studies of the axion-mediated interactions showed that electronic correlations can lead to significant corrections to the value of the molecular parameters. This may increase the sensitivity of $W_{ax}$ to the vibrations of the molecule.

\begin{figure}
    \centering
    \includegraphics[width=0.7\linewidth]{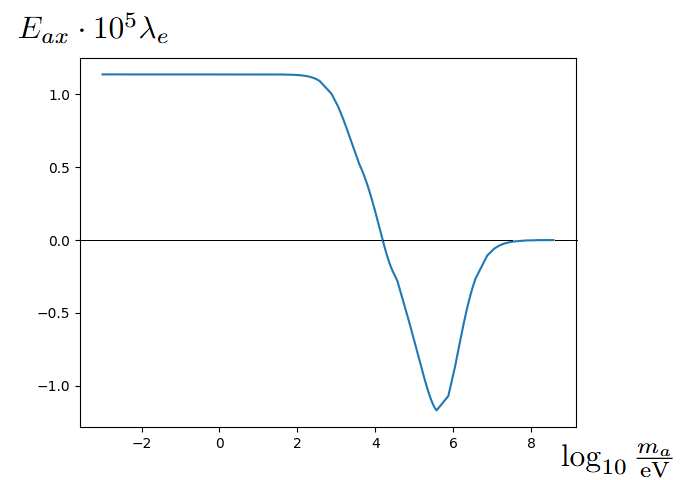}
    \caption{The dependence of $W_{ax}$ on $m_a$}
    \label{fig:m_dependence}
\end{figure}

To test the validity of our computation we have also computed the values for different masses $m_a$ to compare the result with the prediction using the approximation \eqref{WsApprox} and the value for $W_s$ taken from \cite{ourRaOH}. As both properties shown very similar behavior we have restricted ourselves to the equilibrium configurations. The dependence is depicted on Fig. \ref{fig:m_dependence}. For $m_a > 5\cdot 10^7 eV$ the deviation from predictions \eqref{WsApprox} are less than 10\%. Note that just as in \cite{prosnyak2024axion} we see a change of the sign of the property near $m_a\sim 10^{4} {\rm eV}$ which results in the blind spot for this mass range.

\begin{figure}
    \centering
    \includegraphics[width=0.7\linewidth]{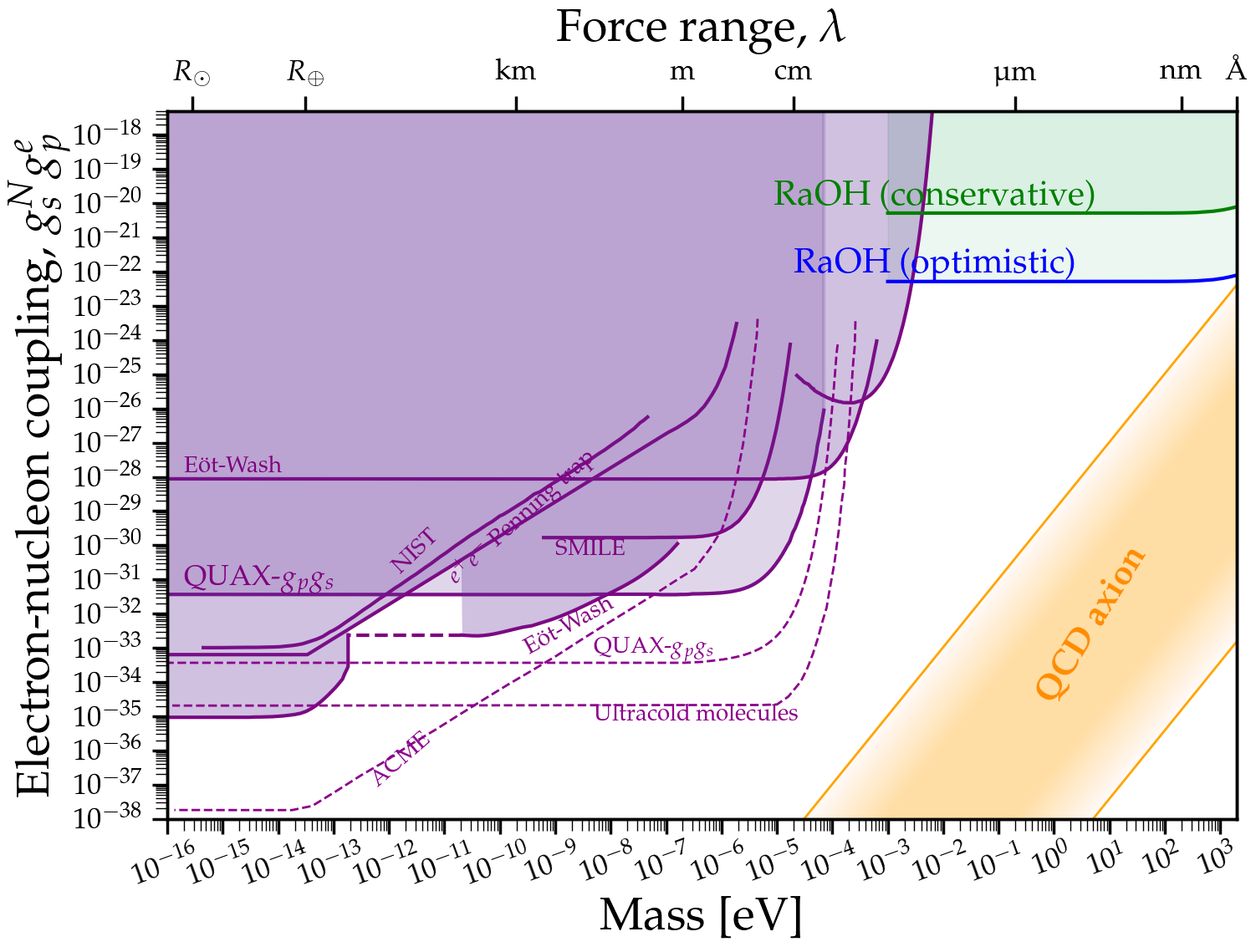}
    \caption{The existing laboratory constraints on the $g_{e,P}g_{N,S}$ and the projected RaOH constraints. The conservative constraint corresponds to the present level of energy split measurement achieved on diatomics, whereas the optimistic ones are given by the expected increase of precision on the triatomic molecules.}
    \label{fig:constraint}
\end{figure}

Finally on Fig. \ref{fig:constraint} we contrast the expected sensitivity of the RaOH experiment with the existing laboratory limits on the product of the axion-nucleon scalar and axion-electron pseudoscalar coupling constants \cite{OHare:2020wah}. Current experiments with diatomic molecules putting a limit $d_e\leq 1.3\cdot 10^{-30}\, e\cdot {\rm cm}$ for $\mathcal{E}=23\,{\rm GV}/{\rm cm}$ are able to sense energy differences $\Delta E = d_e\mathcal{E}\sim 0.7\cdot 10^{-5} Hz$ \cite{roussy2023improved} which gives us the conservative estimate for the RaOH experiment constraint. It is expected \cite{Kozyryev:2017cwq} that for the triatomic molecules this sensitivity may be improved by about two orders of magnitude which yields the optimistic constraint. This is still not enough to beat the astrophysical constraints coming from the red giants $g_{e, P}g_{N, S} < 10^{-26}$, however this would make such molecular experiments best among laboratory experiments for the mass range $m_a > 1\rm{eV}$.

Radium is a heavier atom than Ytterbium, which have resulted in higher values for sensitivities to such interactions as eEDM and eN-SPS for RaOH compared to YbOH molecules \cite{ourRaOH, zakharova2022rotating}. One could expect a higher value for the considered axion-mediated effect in the RaOH molecule, but this is not a case. Hence, while our result shows that the RaOH molecule is less sensitive to this particular interaction, this result nevertheless is interesting. On the other hand, it is important to know such parameters for future experiment. Considering a different ratio with YbOH value of this parameter compared to the eEDM and eN-SPS parameters, performing the measurement with different molecules would allow  to distinguish different P,T-odd effects from each others. While eEDM and eN-SPS are amplified for the RaOH compared to the YbOH by a similar factor, the axion-mediated interaction is significantly reduced and this could facilitate this separation.

\section{Acknowledgement}
The work was supported by the Russian Science Foundation (grant number 24-72-00002).


\begin{thebibliography}{55}
\expandafter\ifx\csname natexlab\endcsname\relax\def\natexlab#1{#1}\fi
\expandafter\ifx\csname bibnamefont\endcsname\relax
  \def\bibnamefont#1{#1}\fi
\expandafter\ifx\csname bibfnamefont\endcsname\relax
  \def\bibfnamefont#1{#1}\fi
\expandafter\ifx\csname citenamefont\endcsname\relax
  \def\citenamefont#1{#1}\fi
\expandafter\ifx\csname url\endcsname\relax
  \def\url#1{\texttt{#1}}\fi
\expandafter\ifx\csname urlprefix\endcsname\relax\def\urlprefix{URL }\fi
\providecommand{\bibinfo}[2]{#2}
\providecommand{\eprint}[2][]{\url{#2}}

\bibitem[{\citenamefont{Chadha-Day et~al.}(2022)\citenamefont{Chadha-Day, Ellis, and Marsh}}]{chadha2022axion}
\bibinfo{author}{\bibfnamefont{F.}~\bibnamefont{Chadha-Day}}, \bibinfo{author}{\bibfnamefont{J.}~\bibnamefont{Ellis}}, \bibnamefont{and} \bibinfo{author}{\bibfnamefont{D.~J.} \bibnamefont{Marsh}}, \bibinfo{journal}{Science advances} \textbf{\bibinfo{volume}{8}}, \bibinfo{pages}{eabj3618} (\bibinfo{year}{2022}).

\bibitem[{\citenamefont{Preskill et~al.}(1983)\citenamefont{Preskill, Wise, and Wilczek}}]{cosmo:1983}
\bibinfo{author}{\bibfnamefont{J.}~\bibnamefont{Preskill}}, \bibinfo{author}{\bibfnamefont{M.~B.} \bibnamefont{Wise}}, \bibnamefont{and} \bibinfo{author}{\bibfnamefont{F.}~\bibnamefont{Wilczek}}, \bibinfo{journal}{Physics Letters B} \textbf{\bibinfo{volume}{120}}, \bibinfo{pages}{127} (\bibinfo{year}{1983}).

\bibitem[{\citenamefont{Abbott and Sikivie}(1983)}]{abbott1983cosmological}
\bibinfo{author}{\bibfnamefont{L.~F.} \bibnamefont{Abbott}} \bibnamefont{and} \bibinfo{author}{\bibfnamefont{P.}~\bibnamefont{Sikivie}}, \bibinfo{journal}{Physics Letters B} \textbf{\bibinfo{volume}{120}}, \bibinfo{pages}{133} (\bibinfo{year}{1983}).

\bibitem[{\citenamefont{Dine and Fischler}(1983)}]{dine1983not}
\bibinfo{author}{\bibfnamefont{M.}~\bibnamefont{Dine}} \bibnamefont{and} \bibinfo{author}{\bibfnamefont{W.}~\bibnamefont{Fischler}}, \bibinfo{journal}{Physics Letters B} \textbf{\bibinfo{volume}{120}}, \bibinfo{pages}{137} (\bibinfo{year}{1983}).

\bibitem[{\citenamefont{Peccei and Quinn}(1977)}]{peccei1977cp}
\bibinfo{author}{\bibfnamefont{R.~D.} \bibnamefont{Peccei}} \bibnamefont{and} \bibinfo{author}{\bibfnamefont{H.~R.} \bibnamefont{Quinn}}, \bibinfo{journal}{Physical Review Letters} \textbf{\bibinfo{volume}{38}}, \bibinfo{pages}{1440} (\bibinfo{year}{1977}).

\bibitem[{\citenamefont{Schwartz}(2014)}]{schwartz2014quantum}
\bibinfo{author}{\bibfnamefont{M.~D.} \bibnamefont{Schwartz}}, \emph{\bibinfo{title}{Quantum field theory and the standard model}} (\bibinfo{publisher}{Cambridge University Press}, \bibinfo{year}{2014}).

\bibitem[{\citenamefont{Navas et~al.}(2024)}]{ParticleDataGroup:2024cfk}
\bibinfo{author}{\bibfnamefont{S.}~\bibnamefont{Navas}} \bibnamefont{et~al.} (\bibinfo{collaboration}{Particle Data Group}), \bibinfo{journal}{Phys. Rev. D} \textbf{\bibinfo{volume}{110}}, \bibinfo{pages}{030001} (\bibinfo{year}{2024}), \urlprefix\url{https://journals.aps.org/prd/abstract/10.1103/PhysRevD.110.030001}.

\bibitem[{\citenamefont{Cabibbo}(1963)}]{Cabibbo1963}
\bibinfo{author}{\bibfnamefont{N.}~\bibnamefont{Cabibbo}}, \bibinfo{journal}{Phys. Rev. Lett.} \textbf{\bibinfo{volume}{10}}, \bibinfo{pages}{531} (\bibinfo{year}{1963}).

\bibitem[{\citenamefont{Kobayashi and Maskawa}(1973)}]{KobayashiMaskawa1973}
\bibinfo{author}{\bibfnamefont{M.}~\bibnamefont{Kobayashi}} \bibnamefont{and} \bibinfo{author}{\bibfnamefont{T.}~\bibnamefont{Maskawa}}, \bibinfo{journal}{Prog. Theor. Phys.} \textbf{\bibinfo{volume}{49}}, \bibinfo{pages}{652} (\bibinfo{year}{1973}).

\bibitem[{\citenamefont{Pontecorvo}(1957)}]{Pontecorvo1957}
\bibinfo{author}{\bibfnamefont{B.}~\bibnamefont{Pontecorvo}}, \bibinfo{journal}{Zh. Eksp. Teor. Fiz.} \textbf{\bibinfo{volume}{34}}, \bibinfo{pages}{247} (\bibinfo{year}{1957}).

\bibitem[{\citenamefont{Maki et~al.}(1962)\citenamefont{Maki, Nakagawa, and Sakata}}]{MNS1962}
\bibinfo{author}{\bibfnamefont{Z.}~\bibnamefont{Maki}}, \bibinfo{author}{\bibfnamefont{M.}~\bibnamefont{Nakagawa}}, \bibnamefont{and} \bibinfo{author}{\bibfnamefont{S.}~\bibnamefont{Sakata}}, \bibinfo{journal}{Prog. Theor. Phys.} \textbf{\bibinfo{volume}{28}}, \bibinfo{pages}{870} (\bibinfo{year}{1962}).

\bibitem[{\citenamefont{FUKUYAMA}(2012)}]{Fukuyama2012}
\bibinfo{author}{\bibfnamefont{T.}~\bibnamefont{FUKUYAMA}}, \bibinfo{journal}{International Journal of Modern Physics A} \textbf{\bibinfo{volume}{27}}, \bibinfo{pages}{1230015} (\bibinfo{year}{2012}), ISSN \bibinfo{issn}{0217-751X}, \urlprefix\url{http://www.worldscientific.com/doi/abs/10.1142/S0217751X12300153}.

\bibitem[{\citenamefont{Pospelov and Ritz}(2014)}]{PospelovRitz2014}
\bibinfo{author}{\bibfnamefont{M.}~\bibnamefont{Pospelov}} \bibnamefont{and} \bibinfo{author}{\bibfnamefont{A.}~\bibnamefont{Ritz}}, \bibinfo{journal}{Phys. Rev. D} \textbf{\bibinfo{volume}{89}}, \bibinfo{pages}{056006} (\bibinfo{year}{2014}), \eprint{1311.5537}.

\bibitem[{\citenamefont{Yamaguchi and Yamanaka}(2020)}]{YamaguchiYamanaka2020}
\bibinfo{author}{\bibfnamefont{Y.}~\bibnamefont{Yamaguchi}} \bibnamefont{and} \bibinfo{author}{\bibfnamefont{N.}~\bibnamefont{Yamanaka}}, \bibinfo{journal}{Phys. Rev. Lett.} \textbf{\bibinfo{volume}{125}}, \bibinfo{pages}{241802} (\bibinfo{year}{2020}), \eprint{2003.08195}.

\bibitem[{\citenamefont{Yamaguchi and Yamanaka}(2021)}]{YamaguchiYamanaka2021}
\bibinfo{author}{\bibfnamefont{Y.}~\bibnamefont{Yamaguchi}} \bibnamefont{and} \bibinfo{author}{\bibfnamefont{N.}~\bibnamefont{Yamanaka}}, \bibinfo{journal}{Phys. Rev. D} \textbf{\bibinfo{volume}{103}}, \bibinfo{pages}{013001} (\bibinfo{year}{2021}), \eprint{2006.00281}.

\bibitem[{\citenamefont{Sakharov}(1998)}]{sakharov1998violation}
\bibinfo{author}{\bibfnamefont{A.~D.} \bibnamefont{Sakharov}}, in \emph{\bibinfo{booktitle}{In The Intermissions… Collected Works on Research into the Essentials of Theoretical Physics in Russian Federal Nuclear Center, Arzamas-16}} (\bibinfo{publisher}{World Scientific}, \bibinfo{year}{1998}), pp. \bibinfo{pages}{84--87}.

\bibitem[{\citenamefont{Ginges and Flambaum}(2004)}]{ginges2004violations}
\bibinfo{author}{\bibfnamefont{J.}~\bibnamefont{Ginges}} \bibnamefont{and} \bibinfo{author}{\bibfnamefont{V.~V.} \bibnamefont{Flambaum}}, \bibinfo{journal}{Physics Reports} \textbf{\bibinfo{volume}{397}}, \bibinfo{pages}{63} (\bibinfo{year}{2004}).

\bibitem[{\citenamefont{Baron et~al.}(2014)\citenamefont{Baron, Campbell, DeMille, Doyle, Gabrielse, Gurevich, Hess, Hutzler, Kirilov, Kozyryev et~al.}}]{baron2014order}
\bibinfo{author}{\bibfnamefont{J.}~\bibnamefont{Baron}}, \bibinfo{author}{\bibfnamefont{W.~C.} \bibnamefont{Campbell}}, \bibinfo{author}{\bibfnamefont{D.}~\bibnamefont{DeMille}}, \bibinfo{author}{\bibfnamefont{J.~M.} \bibnamefont{Doyle}}, \bibinfo{author}{\bibfnamefont{G.}~\bibnamefont{Gabrielse}}, \bibinfo{author}{\bibfnamefont{Y.~V.} \bibnamefont{Gurevich}}, \bibinfo{author}{\bibfnamefont{P.~W.} \bibnamefont{Hess}}, \bibinfo{author}{\bibfnamefont{N.~R.} \bibnamefont{Hutzler}}, \bibinfo{author}{\bibfnamefont{E.}~\bibnamefont{Kirilov}}, \bibinfo{author}{\bibfnamefont{I.}~\bibnamefont{Kozyryev}}, \bibnamefont{et~al.}, \bibinfo{journal}{Science} \textbf{\bibinfo{volume}{343}}, \bibinfo{pages}{269} (\bibinfo{year}{2014}).

\bibitem[{\citenamefont{Chubukov et~al.}(2019)\citenamefont{Chubukov, Skripnikov, and Labzowsky}}]{ChubukovLabzowsky2016}
\bibinfo{author}{\bibfnamefont{D.~V.} \bibnamefont{Chubukov}}, \bibinfo{author}{\bibfnamefont{L.~V.} \bibnamefont{Skripnikov}}, \bibnamefont{and} \bibinfo{author}{\bibfnamefont{L.~N.} \bibnamefont{Labzowsky}}, \bibinfo{journal}{JETP Lett.} \textbf{\bibinfo{volume}{110}}, \bibinfo{pages}{382} (\bibinfo{year}{2019}).

\bibitem[{\citenamefont{Andreev et~al.}(2018)\citenamefont{Andreev, Ang, DeMille, Doyle, Gabrielse, Haefner, Hutzler, Lasner, Meisenhelder, O'Leary et~al.}}]{ACME:18}
\bibinfo{author}{\bibfnamefont{V.}~\bibnamefont{Andreev}}, \bibinfo{author}{\bibfnamefont{D.}~\bibnamefont{Ang}}, \bibinfo{author}{\bibfnamefont{D.}~\bibnamefont{DeMille}}, \bibinfo{author}{\bibfnamefont{J.}~\bibnamefont{Doyle}}, \bibinfo{author}{\bibfnamefont{G.}~\bibnamefont{Gabrielse}}, \bibinfo{author}{\bibfnamefont{J.}~\bibnamefont{Haefner}}, \bibinfo{author}{\bibfnamefont{N.}~\bibnamefont{Hutzler}}, \bibinfo{author}{\bibfnamefont{Z.}~\bibnamefont{Lasner}}, \bibinfo{author}{\bibfnamefont{C.}~\bibnamefont{Meisenhelder}}, \bibinfo{author}{\bibfnamefont{B.}~\bibnamefont{O'Leary}}, \bibnamefont{et~al.}, \bibinfo{journal}{Nature} \textbf{\bibinfo{volume}{562}}, \bibinfo{pages}{355} (\bibinfo{year}{2018}).

\bibitem[{\citenamefont{Roussy et~al.}(2023)\citenamefont{Roussy, Caldwell, Wright, Cairncross, Shagam, Ng, Schlossberger, Park, Wang, Ye et~al.}}]{roussy2023improved}
\bibinfo{author}{\bibfnamefont{T.~S.} \bibnamefont{Roussy}}, \bibinfo{author}{\bibfnamefont{L.}~\bibnamefont{Caldwell}}, \bibinfo{author}{\bibfnamefont{T.}~\bibnamefont{Wright}}, \bibinfo{author}{\bibfnamefont{W.~B.} \bibnamefont{Cairncross}}, \bibinfo{author}{\bibfnamefont{Y.}~\bibnamefont{Shagam}}, \bibinfo{author}{\bibfnamefont{K.~B.} \bibnamefont{Ng}}, \bibinfo{author}{\bibfnamefont{N.}~\bibnamefont{Schlossberger}}, \bibinfo{author}{\bibfnamefont{S.~Y.} \bibnamefont{Park}}, \bibinfo{author}{\bibfnamefont{A.}~\bibnamefont{Wang}}, \bibinfo{author}{\bibfnamefont{J.}~\bibnamefont{Ye}}, \bibnamefont{et~al.}, \bibinfo{journal}{Science} \textbf{\bibinfo{volume}{381}}, \bibinfo{pages}{46} (\bibinfo{year}{2023}).

\bibitem[{\citenamefont{Kozlov and Labzowsky}(1995)}]{KozlovLabzowsky1995}
\bibinfo{author}{\bibfnamefont{M.~G.} \bibnamefont{Kozlov}} \bibnamefont{and} \bibinfo{author}{\bibfnamefont{L.~N.} \bibnamefont{Labzowsky}}, \bibinfo{journal}{Journal of Physics B: Atomic, Molecular and Optical Physics} \textbf{\bibinfo{volume}{28}}, \bibinfo{pages}{1933} (\bibinfo{year}{1995}).

\bibitem[{\citenamefont{Gaul et~al.}(2020)\citenamefont{Gaul, Kozlov, Isaev, and Berger}}]{gaul2020chiral}
\bibinfo{author}{\bibfnamefont{K.}~\bibnamefont{Gaul}}, \bibinfo{author}{\bibfnamefont{M.~G.} \bibnamefont{Kozlov}}, \bibinfo{author}{\bibfnamefont{T.~A.} \bibnamefont{Isaev}}, \bibnamefont{and} \bibinfo{author}{\bibfnamefont{R.}~\bibnamefont{Berger}}, \bibinfo{journal}{Physical review letters} \textbf{\bibinfo{volume}{125}}, \bibinfo{pages}{123004} (\bibinfo{year}{2020}).

\bibitem[{\citenamefont{Prosnyak et~al.}(2023)\citenamefont{Prosnyak, Maison, and Skripnikov}}]{Prosnyak:2023duq}
\bibinfo{author}{\bibfnamefont{S.~D.} \bibnamefont{Prosnyak}}, \bibinfo{author}{\bibfnamefont{D.~E.} \bibnamefont{Maison}}, \bibnamefont{and} \bibinfo{author}{\bibfnamefont{L.~V.} \bibnamefont{Skripnikov}}, \bibinfo{journal}{Symmetry} \textbf{\bibinfo{volume}{15}}, \bibinfo{pages}{1043} (\bibinfo{year}{2023}), \eprint{2304.07164}.

\bibitem[{\citenamefont{Prosnyak and Skripnikov}(2024)}]{prosnyak2024axion}
\bibinfo{author}{\bibfnamefont{S.~D.} \bibnamefont{Prosnyak}} \bibnamefont{and} \bibinfo{author}{\bibfnamefont{L.~V.} \bibnamefont{Skripnikov}}, \bibinfo{journal}{Physical Review A} \textbf{\bibinfo{volume}{109}}, \bibinfo{pages}{042821} (\bibinfo{year}{2024}).

\bibitem[{\citenamefont{Maison et~al.}(2021{\natexlab{a}})\citenamefont{Maison, Flambaum, Hutzler, and Skripnikov}}]{maison2020study}
\bibinfo{author}{\bibfnamefont{D.~E.} \bibnamefont{Maison}}, \bibinfo{author}{\bibfnamefont{V.~V.} \bibnamefont{Flambaum}}, \bibinfo{author}{\bibfnamefont{N.~R.} \bibnamefont{Hutzler}}, \bibnamefont{and} \bibinfo{author}{\bibfnamefont{L.~V.} \bibnamefont{Skripnikov}}, \bibinfo{journal}{Phys. Rev. A} \textbf{\bibinfo{volume}{103}}, \bibinfo{pages}{022813} (\bibinfo{year}{2021}{\natexlab{a}}), \eprint{2010.11669}.

\bibitem[{\citenamefont{Maison et~al.}(2021{\natexlab{b}})\citenamefont{Maison, Skripnikov, Oleynichenko, and Zaitsevskii}}]{maison2021axion}
\bibinfo{author}{\bibfnamefont{D.}~\bibnamefont{Maison}}, \bibinfo{author}{\bibfnamefont{L.}~\bibnamefont{Skripnikov}}, \bibinfo{author}{\bibfnamefont{A.}~\bibnamefont{Oleynichenko}}, \bibnamefont{and} \bibinfo{author}{\bibfnamefont{A.}~\bibnamefont{Zaitsevskii}}, \bibinfo{journal}{The Journal of Chemical Physics} \textbf{\bibinfo{volume}{154}}, \bibinfo{pages}{224303} (\bibinfo{year}{2021}{\natexlab{b}}).

\bibitem[{\citenamefont{Stadnik et~al.}(2018)\citenamefont{Stadnik, Dzuba, and Flambaum}}]{stadnik2018improved}
\bibinfo{author}{\bibfnamefont{Y.}~\bibnamefont{Stadnik}}, \bibinfo{author}{\bibfnamefont{V.}~\bibnamefont{Dzuba}}, \bibnamefont{and} \bibinfo{author}{\bibfnamefont{V.}~\bibnamefont{Flambaum}}, \bibinfo{journal}{Physical review letters} \textbf{\bibinfo{volume}{120}}, \bibinfo{pages}{013202} (\bibinfo{year}{2018}).

\bibitem[{\citenamefont{Dzuba et~al.}(2018)\citenamefont{Dzuba, Flambaum, Samsonov, and Stadnik}}]{dzuba2018new}
\bibinfo{author}{\bibfnamefont{V.}~\bibnamefont{Dzuba}}, \bibinfo{author}{\bibfnamefont{V.}~\bibnamefont{Flambaum}}, \bibinfo{author}{\bibfnamefont{I.}~\bibnamefont{Samsonov}}, \bibnamefont{and} \bibinfo{author}{\bibfnamefont{Y.}~\bibnamefont{Stadnik}}, \bibinfo{journal}{Physical Review D} \textbf{\bibinfo{volume}{98}}, \bibinfo{pages}{035048} (\bibinfo{year}{2018}).

\bibitem[{\citenamefont{Zakharova and Petrov}(2021)}]{ourRaOH}
\bibinfo{author}{\bibfnamefont{A.}~\bibnamefont{Zakharova}} \bibnamefont{and} \bibinfo{author}{\bibfnamefont{A.}~\bibnamefont{Petrov}}, \bibinfo{journal}{Phys. Rev. A} \textbf{\bibinfo{volume}{103}}, \bibinfo{pages}{032819} (\bibinfo{year}{2021}), \eprint{2012.08427}.

\bibitem[{\citenamefont{Zakharova et~al.}(2021)\citenamefont{Zakharova, Kurchavov, and Petrov}}]{zakharova2021rovibrational}
\bibinfo{author}{\bibfnamefont{A.}~\bibnamefont{Zakharova}}, \bibinfo{author}{\bibfnamefont{I.}~\bibnamefont{Kurchavov}}, \bibnamefont{and} \bibinfo{author}{\bibfnamefont{A.}~\bibnamefont{Petrov}}, \bibinfo{journal}{The Journal of Chemical Physics} \textbf{\bibinfo{volume}{155}}, \bibinfo{pages}{164301} (\bibinfo{year}{2021}).

\bibitem[{\citenamefont{Zakharova and Petrov}(2022)}]{zakharova2022impact}
\bibinfo{author}{\bibfnamefont{A.}~\bibnamefont{Zakharova}} \bibnamefont{and} \bibinfo{author}{\bibfnamefont{A.}~\bibnamefont{Petrov}}, \bibinfo{journal}{The Journal of Chemical Physics} \textbf{\bibinfo{volume}{157}} (\bibinfo{year}{2022}).

\bibitem[{\citenamefont{Petrov and Zakharova}(2022)}]{petrov2022sensitivity}
\bibinfo{author}{\bibfnamefont{A.}~\bibnamefont{Petrov}} \bibnamefont{and} \bibinfo{author}{\bibfnamefont{A.}~\bibnamefont{Zakharova}}, \bibinfo{journal}{Physical Review A} \textbf{\bibinfo{volume}{105}}, \bibinfo{pages}{L050801} (\bibinfo{year}{2022}).

\bibitem[{\citenamefont{Zakharova}(2022)}]{zakharova2022rotating}
\bibinfo{author}{\bibfnamefont{A.}~\bibnamefont{Zakharova}}, \bibinfo{journal}{Physical Review A} \textbf{\bibinfo{volume}{105}}, \bibinfo{pages}{032811} (\bibinfo{year}{2022}).

\bibitem[{\citenamefont{Zakharova}(2024)}]{zakharova2024symmetric}
\bibinfo{author}{\bibfnamefont{A.}~\bibnamefont{Zakharova}}, \bibinfo{journal}{Chemical Physics Letters} \textbf{\bibinfo{volume}{854}}, \bibinfo{pages}{141552} (\bibinfo{year}{2024}).

\bibitem[{\citenamefont{Zakharova}(2025{\natexlab{a}})}]{zakharova2025hexatomic}
\bibinfo{author}{\bibfnamefont{A.~V.} \bibnamefont{Zakharova}}, \bibinfo{journal}{Optics and Spectroscopy} \textbf{\bibinfo{volume}{133}}, \bibinfo{pages}{697} (\bibinfo{year}{2025}{\natexlab{a}}).

\bibitem[{\citenamefont{Isaev and Berger}(2016)}]{isaev2016polyatomic}
\bibinfo{author}{\bibfnamefont{T.~A.} \bibnamefont{Isaev}} \bibnamefont{and} \bibinfo{author}{\bibfnamefont{R.}~\bibnamefont{Berger}}, \bibinfo{journal}{Physical review letters} \textbf{\bibinfo{volume}{116}}, \bibinfo{pages}{063006} (\bibinfo{year}{2016}).

\bibitem[{\citenamefont{Kozyryev et~al.}(2016)\citenamefont{Kozyryev, Baum, Matsuda, and Doyle}}]{kozyryev2016proposal}
\bibinfo{author}{\bibfnamefont{I.}~\bibnamefont{Kozyryev}}, \bibinfo{author}{\bibfnamefont{L.}~\bibnamefont{Baum}}, \bibinfo{author}{\bibfnamefont{K.}~\bibnamefont{Matsuda}}, \bibnamefont{and} \bibinfo{author}{\bibfnamefont{J.~M.} \bibnamefont{Doyle}}, \bibinfo{journal}{ChemPhysChem} \textbf{\bibinfo{volume}{17}}, \bibinfo{pages}{3641} (\bibinfo{year}{2016}).

\bibitem[{\citenamefont{Kozyryev et~al.}(2019)\citenamefont{Kozyryev, Steimle, Yu, Nguyen, and Doyle}}]{kozyryev2019determination}
\bibinfo{author}{\bibfnamefont{I.}~\bibnamefont{Kozyryev}}, \bibinfo{author}{\bibfnamefont{T.~C.} \bibnamefont{Steimle}}, \bibinfo{author}{\bibfnamefont{P.}~\bibnamefont{Yu}}, \bibinfo{author}{\bibfnamefont{D.-T.} \bibnamefont{Nguyen}}, \bibnamefont{and} \bibinfo{author}{\bibfnamefont{J.~M.} \bibnamefont{Doyle}}, \bibinfo{journal}{New Journal of Physics} \textbf{\bibinfo{volume}{21}}, \bibinfo{pages}{052002} (\bibinfo{year}{2019}).

\bibitem[{\citenamefont{Augenbraun et~al.}(2021)\citenamefont{Augenbraun, Lasner, Frenett, Sawaoka, Le, Doyle, and Steimle}}]{augenbraun2021observation}
\bibinfo{author}{\bibfnamefont{B.~L.} \bibnamefont{Augenbraun}}, \bibinfo{author}{\bibfnamefont{Z.~D.} \bibnamefont{Lasner}}, \bibinfo{author}{\bibfnamefont{A.}~\bibnamefont{Frenett}}, \bibinfo{author}{\bibfnamefont{H.}~\bibnamefont{Sawaoka}}, \bibinfo{author}{\bibfnamefont{A.~T.} \bibnamefont{Le}}, \bibinfo{author}{\bibfnamefont{J.~M.} \bibnamefont{Doyle}}, \bibnamefont{and} \bibinfo{author}{\bibfnamefont{T.~C.} \bibnamefont{Steimle}}, \bibinfo{journal}{Physical Review A} \textbf{\bibinfo{volume}{103}}, \bibinfo{pages}{022814} (\bibinfo{year}{2021}).

\bibitem[{\citenamefont{Mitra et~al.}(2020)\citenamefont{Mitra, Vilas, Hallas, Anderegg, Augenbraun, Baum, Miller, Raval, and Doyle}}]{mitra2020direct}
\bibinfo{author}{\bibfnamefont{D.}~\bibnamefont{Mitra}}, \bibinfo{author}{\bibfnamefont{N.~B.} \bibnamefont{Vilas}}, \bibinfo{author}{\bibfnamefont{C.}~\bibnamefont{Hallas}}, \bibinfo{author}{\bibfnamefont{L.}~\bibnamefont{Anderegg}}, \bibinfo{author}{\bibfnamefont{B.~L.} \bibnamefont{Augenbraun}}, \bibinfo{author}{\bibfnamefont{L.}~\bibnamefont{Baum}}, \bibinfo{author}{\bibfnamefont{C.}~\bibnamefont{Miller}}, \bibinfo{author}{\bibfnamefont{S.}~\bibnamefont{Raval}}, \bibnamefont{and} \bibinfo{author}{\bibfnamefont{J.~M.} \bibnamefont{Doyle}}, \bibinfo{journal}{Science} \textbf{\bibinfo{volume}{369}}, \bibinfo{pages}{1366} (\bibinfo{year}{2020}).

\bibitem[{\citenamefont{Fan et~al.}(2021)\citenamefont{Fan, Holliman, Shi, Zhang, Straus, Li, Buechele, and Jayich}}]{fan2021optical}
\bibinfo{author}{\bibfnamefont{M.}~\bibnamefont{Fan}}, \bibinfo{author}{\bibfnamefont{C.}~\bibnamefont{Holliman}}, \bibinfo{author}{\bibfnamefont{X.}~\bibnamefont{Shi}}, \bibinfo{author}{\bibfnamefont{H.}~\bibnamefont{Zhang}}, \bibinfo{author}{\bibfnamefont{M.}~\bibnamefont{Straus}}, \bibinfo{author}{\bibfnamefont{X.}~\bibnamefont{Li}}, \bibinfo{author}{\bibfnamefont{S.}~\bibnamefont{Buechele}}, \bibnamefont{and} \bibinfo{author}{\bibfnamefont{A.}~\bibnamefont{Jayich}}, \bibinfo{journal}{Physical Review Letters} \textbf{\bibinfo{volume}{126}}, \bibinfo{pages}{023002} (\bibinfo{year}{2021}).

\bibitem[{\citenamefont{Sawaoka et~al.}(2025)\citenamefont{Sawaoka, Nasir, Lunstad, Li, Mango, Lasner, and Doyle}}]{sawaoka2025optical}
\bibinfo{author}{\bibfnamefont{H.}~\bibnamefont{Sawaoka}}, \bibinfo{author}{\bibfnamefont{A.}~\bibnamefont{Nasir}}, \bibinfo{author}{\bibfnamefont{A.}~\bibnamefont{Lunstad}}, \bibinfo{author}{\bibfnamefont{M.}~\bibnamefont{Li}}, \bibinfo{author}{\bibfnamefont{J.}~\bibnamefont{Mango}}, \bibinfo{author}{\bibfnamefont{Z.~D.} \bibnamefont{Lasner}}, \bibnamefont{and} \bibinfo{author}{\bibfnamefont{J.~M.} \bibnamefont{Doyle}}, \bibinfo{journal}{arXiv preprint arXiv:2509.01618}  (\bibinfo{year}{2025}).

\bibitem[{\citenamefont{Conn et~al.}(2025)\citenamefont{Conn, Yu, Howard, Yang, Zhang, Jadbabaie, Gorou, Gaiser, Steimle, Cheng et~al.}}]{conn2025production}
\bibinfo{author}{\bibfnamefont{C.~J.} \bibnamefont{Conn}}, \bibinfo{author}{\bibfnamefont{P.}~\bibnamefont{Yu}}, \bibinfo{author}{\bibfnamefont{M.~I.} \bibnamefont{Howard}}, \bibinfo{author}{\bibfnamefont{Y.}~\bibnamefont{Yang}}, \bibinfo{author}{\bibfnamefont{C.}~\bibnamefont{Zhang}}, \bibinfo{author}{\bibfnamefont{A.}~\bibnamefont{Jadbabaie}}, \bibinfo{author}{\bibfnamefont{A.}~\bibnamefont{Gorou}}, \bibinfo{author}{\bibfnamefont{A.~N.} \bibnamefont{Gaiser}}, \bibinfo{author}{\bibfnamefont{T.~C.} \bibnamefont{Steimle}}, \bibinfo{author}{\bibfnamefont{L.}~\bibnamefont{Cheng}}, \bibnamefont{et~al.}, \bibinfo{journal}{arXiv preprint arXiv:2508.08368}  (\bibinfo{year}{2025}).

\bibitem[{\citenamefont{Titov and Mosyagin}(1999)}]{titov1999generalized}
\bibinfo{author}{\bibfnamefont{A.}~\bibnamefont{Titov}} \bibnamefont{and} \bibinfo{author}{\bibfnamefont{N.}~\bibnamefont{Mosyagin}}, \bibinfo{journal}{International journal of quantum chemistry} \textbf{\bibinfo{volume}{71}}, \bibinfo{pages}{359} (\bibinfo{year}{1999}).

\bibitem[{\citenamefont{Mosyagin et~al.}(2010)\citenamefont{Mosyagin, Zaitsevskii, and Titov}}]{mosyagin2010shape}
\bibinfo{author}{\bibfnamefont{N.~S.} \bibnamefont{Mosyagin}}, \bibinfo{author}{\bibfnamefont{A.}~\bibnamefont{Zaitsevskii}}, \bibnamefont{and} \bibinfo{author}{\bibfnamefont{A.~V.} \bibnamefont{Titov}}, \bibinfo{journal}{International Review of Atomic and Molecular Physics} \textbf{\bibinfo{volume}{1}}, \bibinfo{pages}{63} (\bibinfo{year}{2010}).

\bibitem[{\citenamefont{Mosyagin et~al.}(2016)\citenamefont{Mosyagin, Zaitsevskii, Skripnikov, and Titov}}]{mosyagin2016generalized}
\bibinfo{author}{\bibfnamefont{N.~S.} \bibnamefont{Mosyagin}}, \bibinfo{author}{\bibfnamefont{A.~V.} \bibnamefont{Zaitsevskii}}, \bibinfo{author}{\bibfnamefont{L.~V.} \bibnamefont{Skripnikov}}, \bibnamefont{and} \bibinfo{author}{\bibfnamefont{A.~V.} \bibnamefont{Titov}}, \bibinfo{journal}{International Journal of Quantum Chemistry} \textbf{\bibinfo{volume}{116}}, \bibinfo{pages}{301} (\bibinfo{year}{2016}).

\bibitem[{\citenamefont{Titov et~al.}(2006)\citenamefont{Titov, Mosyagin, Petrov, Isaev, and DeMille}}]{titov2006d}
\bibinfo{author}{\bibfnamefont{A.}~\bibnamefont{Titov}}, \bibinfo{author}{\bibfnamefont{N.}~\bibnamefont{Mosyagin}}, \bibinfo{author}{\bibfnamefont{A.}~\bibnamefont{Petrov}}, \bibinfo{author}{\bibfnamefont{T.}~\bibnamefont{Isaev}}, \bibnamefont{and} \bibinfo{author}{\bibfnamefont{D.}~\bibnamefont{DeMille}}, in \emph{\bibinfo{booktitle}{Recent Advances in the Theory of Chemical and Physical Systems}} (\bibinfo{publisher}{Springer}, \bibinfo{year}{2006}), pp. \bibinfo{pages}{253--283}.

\bibitem[{\citenamefont{Zakharova}(2025{\natexlab{b}})}]{zakharova2025parallelized}
\bibinfo{author}{\bibfnamefont{A.}~\bibnamefont{Zakharova}}, in \emph{\bibinfo{booktitle}{Russian Supercomputing Days}} (\bibinfo{publisher}{Springer}, \bibinfo{year}{2025}{\natexlab{b}}), pp. \bibinfo{pages}{214--225}.

\bibitem[{\citenamefont{Huber and Herzberg}(1979)}]{Huber:1979}
\bibinfo{author}{\bibfnamefont{K.~P.} \bibnamefont{Huber}} \bibnamefont{and} \bibinfo{author}{\bibfnamefont{G.}~\bibnamefont{Herzberg}}, \emph{\bibinfo{title}{Constants of Diatomic Molecules}} (\bibinfo{publisher}{Van Nostrand-Reinhold}, \bibinfo{address}{New York}, \bibinfo{year}{1979}).

\bibitem[{\citenamefont{McGuire and Kouri}(1974)}]{mcguire1974quantum}
\bibinfo{author}{\bibfnamefont{P.}~\bibnamefont{McGuire}} \bibnamefont{and} \bibinfo{author}{\bibfnamefont{D.~J.} \bibnamefont{Kouri}}, \bibinfo{journal}{The Journal of Chemical Physics} \textbf{\bibinfo{volume}{60}}, \bibinfo{pages}{2488} (\bibinfo{year}{1974}).

\bibitem[{\citenamefont{{URL: http://www.qchem.pnpi.spb.ru/Basis/~}}()}]{QCPNPI:Basis}
\bibinfo{author}{\bibnamefont{{URL: http://www.qchem.pnpi.spb.ru/Basis/~}}}, \bibinfo{note}{~{GRECPs} and basis sets}.

\bibitem[{\citenamefont{Titov et~al.}(2003)\citenamefont{Titov, Mosyagin, Isaev, and Petrov}}]{titov2003accuracy}
\bibinfo{author}{\bibfnamefont{A.}~\bibnamefont{Titov}}, \bibinfo{author}{\bibfnamefont{N.}~\bibnamefont{Mosyagin}}, \bibinfo{author}{\bibfnamefont{T.}~\bibnamefont{Isaev}}, \bibnamefont{and} \bibinfo{author}{\bibfnamefont{A.}~\bibnamefont{Petrov}}, \bibinfo{journal}{Physics of Atomic Nuclei} \textbf{\bibinfo{volume}{66}}, \bibinfo{pages}{1152} (\bibinfo{year}{2003}).

\bibitem[{\citenamefont{O'Hare and Vitagliano}(2020)}]{OHare:2020wah}
\bibinfo{author}{\bibfnamefont{C.~A.~J.} \bibnamefont{O'Hare}} \bibnamefont{and} \bibinfo{author}{\bibfnamefont{E.}~\bibnamefont{Vitagliano}}, \bibinfo{journal}{Phys. Rev. D} \textbf{\bibinfo{volume}{102}}, \bibinfo{pages}{115026} (\bibinfo{year}{2020}), \eprint{2010.03889}.

\bibitem[{\citenamefont{Kozyryev and Hutzler}(2017)}]{Kozyryev:2017cwq}
\bibinfo{author}{\bibfnamefont{I.}~\bibnamefont{Kozyryev}} \bibnamefont{and} \bibinfo{author}{\bibfnamefont{N.~R.} \bibnamefont{Hutzler}}, \bibinfo{journal}{Phys. Rev. Lett.} \textbf{\bibinfo{volume}{119}}, \bibinfo{pages}{133002} (\bibinfo{year}{2017}), \eprint{1705.11020}.

\end{thebibliography}
\end{document}